\documentstyle[12pt,epsfig]{article}
\def\f {\beta_f}
\def\a {\beta_a}
\def\v {\beta_v}
\def\b {\beta}
\def\bc {\beta_c}
\def\l {\langle L_a \rangle}
\def\lf {\langle L_f \rangle}
\def\fa {4^3 \times 4}
\def\sa {6^3 \times 4}
\def\ea {8^3 \times 4}
\def\fs {4^4}
\def\xs {6^4}
\def\es {8^4}
\def\tw {8^3 \times 2}

\begin{document}
\begin{titlepage}
\vskip3cm
\begin{flushright}
{\sf TIFR/TH/97-45 \\
ZIF-MS-12/97 \\
hep-lat/9708026}
\end{flushright}
\vskip2cm
\begin{center}
{\Large {\bf Phase Transitions in $SO(3)$ Lattice Gauge Theory}} \\[5mm]
{\large Saumen Datta$^{1,}$\footnote[3]{E-mail:saumen@theory.tifr.res.in} and
Rajiv V. Gavai$^{1,2,}$\footnote[4]{E-mail:gavai@theory.tifr.res.in}} \\[3mm]
{\em ${}^1$Theoretical Physics Group, \\ 
Tata Institute of Fundamental Research, \\
     Homi Bhabha Road, Mumbai-400005, India. \\[5mm]
${}^2$Zentrum f\"ur interdisziplin\"are  Forschung, \\
Wellenberg 1, D-33615 Bielefeld, Germany.} \\[2cm]
\end{center}
\begin{abstract} 
The phase diagram of $SO(3)$ lattice gauge theory is investigated by 
Monte Carlo techniques on both symmetric $N_\sigma^4$ and asymmetric
$N_\sigma^3 \times N_\beta$ lattices with a view (i) to understanding
the relationship between the bulk transition and the deconfinement
transition, and (ii) to resolving the current ambiguity about the nature
of the high temperature phase. A number of tests, including an
introduction of a magnetic field and measurement of different
correlation functions in the phases with positive and negative 
values for the adjoint Polyakov line, $L_a$, lead to the conclusion that the 
two phases correspond to the same physical state. Studies on lattices of
different sizes reveal only one phase transition for
this theory on all of them and it appears to have a deconfining nature.


\end{abstract}
\end{titlepage}

\section{\label{sc1} Introduction}

The formulation of gauge theories on discrete space-time lattices \cite{wil}
provide an elegant way to investigate confinement in non-Abelian gauge
theories.  Using numerical Monte Carlo techniques, it was
shown \cite{cre} that confinement survives the approach to the continuum
limit of $a \rightarrow 0$, where $a$ is the lattice spacing.  The same
techniques enabled one to explore these theories at nonzero temperatures,
where it was found \cite{gav} that both $SU(2)$ and $SU(3)$ Yang-Mills
theories undergo deconfinement phase transitions to a new phase of
deconfined glue.  Exploiting the symmetry of their order parameters, it
was argued \cite{svet} that the $SU(3)$ theory should have a first order
transition, while the critical exponents of the $SU(2)$ theory should be
the same as those of the 3-dimensional Ising model, which was confirmed by 
high precision determination \cite{eng} of the exponents.

Since the continuum limit is at the critical point of a lattice
theory, a large class of actions, which are in the same universality class as
the popular Wilson action \cite{wil}, given by Eq.(\ref{WIL}), 
\begin{equation}
S=\f \sum_p \Bigl[ 1 - {1 \over N} {\rm Re} ~ {\rm Tr}_f  ~ U_p 
\Bigr] ~~,~~\label{WIL} 
\end{equation}
are expected to give rise to the same predictions for continuum physics.
In particular, the trace, taken in the fundamental representation of the 
gauge group in Eq.(\ref{WIL}), can be taken in any  
representation of the gauge group.  Indeed, as the 2-loop $\beta$-function 
for pure $SU(2)$ gauge theory is identical to that of the
pure $SO(3)$ gauge theory, one expects the latter to yield the same
continuum physics.  On the other hand, $SO(3)$ does not
have the $Z(2)$ center symmetry, whose spontaneous breakdown in the $SU(2)$
theory indicates the deconfinement transition.  This makes the investigation of
the phase diagram of the $SO(3)$ gauge theory especially interesting and
important.  It has been argued \cite{smi} that the deconfinement
transition, in this case, will show up
as a crossover which sharpens in the continuum limit to give the
Ising-like second order phase transition.

Another reason for investigating the finite temperature transition in
$SO(3)$ gauge theory is that it is supposed \cite{bha} to have a bulk phase
transition and may thus provide a test case for studying the interplay
between the types of phase transitions.  Recently, simulations of the
Bhanot-Creutz action \cite{bha} for $SU(2)$ gauge theory,
\begin{equation}
S=\sum_p \Bigl[ \f \Bigl( 1 - {1 \over 2} {\rm Tr}_f ~ U_p \Bigr) 
+ \a \Bigl( 1 - {1 \over 3} {\rm Tr}_a ~ U_p \Bigr) \Bigr] 
\label{BHA}
\end{equation}
at finite temperature revealed \cite{gav1} that the known deconfinement 
transition point in Wilson action becomes a line in the $\f - \a$
plane and joins the bulk transition line seen in \cite{bha}. The order
of the deconfinement transition was also seen to change from second to
first for $ \a \ge 1.25 $. At no $\a$, two separate transitions were
found in spite of variations in the lattice size in temporal directions
from $N_t = 2$ to 8. Considering the different physical nature of
these transitions, their coincidence was puzzling.  In view of the
behavior of the order parameter for the deconfinement phase transition, 
it was concluded in \cite{gav2} that the transition seen in \cite{bha} is 
a deconfinement transition rather than a bulk one.  However, very little
shift in the transition coupling was seen under a change of $N_t$,
which is more characteristic of a bulk transition.

The studies in \cite{gav1,gav2} were all done for a relatively small $\a$, 
i.e., close to the Wilson action. In this paper we study the Bhanot-Creutz 
action with $\f = 0$ with an aim to study the issue of
bulk vs deconfinement transitions away from the Wilson action axis. 
As the trace is then taken only in the adjoint representation, it corresponds 
to $SO(3)$ gauge theory. In the strong coupling domain, $SO(3)$ gauge
theory is qualitatively different from $SU(2)$. Its approach to the
continuum theory of su(2) algebra has been studied by Halliday and Schwimmer 
(\ref{HAL}), using a modified action which is similar to $SO(3)$
Wilson action, but which reveals the topological properties explicitly.
They found a phase transition, driven by the melting of the condensate 
of $Z(N)$ monopoles, separating the strong coupling region from the weak 
coupling region.
  
A study of finite temperature $SO(3)$ gauge 
theory was carried out in \cite{sri} and a deconfining transition for
this theory was found.  However, there was some ambiguity about the nature 
of the high temperature phase and the order of the phase transition
in \cite{sri}. In this work, we attempt to clarify the 
nature of the high temperature phase, and the order of the
phase transition for $SO(3)$ lattice gauge theory.

The plan of our paper is as follows: in Sec. \ref{sc2}, we define the
actions and the different observables we use for our study.
In Sec. \ref{sc3}, we discuss the nature of the high temperature
phase with a view to clarify some of the issues in \cite{sri}. 
Finite size scaling analysis is used in the next section to
establish the order of the transition of $SO(3)$ gauge theory. 
In Sec. \ref{sc5}, the issue of
bulk versus deconfinement transition is discussed. The last section
contains a summary of our results and their discussion.

\section{\label{sc2} Actions and Observables}

The Wilson action for $SO(3)$ gauge theory is
\begin{equation}
S = \b \sum_p \Bigl( 1 - {1 \over 3} {\rm Tr} ~ U_p \Bigr) \label{ADJ}
\end{equation}
where $U_p$ denotes the directed product of the basic link variables
which describe the gauge fields, $U_\mu (x) \in SO(3)$, around an
elementary plaquette $p$. Comparing the naive classical continuum limit of Eq.
(\ref{ADJ}) with the standard action for $SU(2)$ Yang-Mills theory, one
obtains $ \b = 3 / 2 g_0^2$, where $g_0$ is the bare coupling constant of the
continuum theory. 

Using the property of the adjoint trace, 
${\rm Tr}_a V = ( {\rm Tr}_f V)^2 -1$, 
the action (\ref{BHA}) can be written for $\f =0$ as 
\begin{equation}
S = {4 \a \over 3} \sum_p \Bigl[ 1 + \Bigl( {1 \over 2} {\rm Tr}_f
~ U_p \Bigr) ^2 \Bigr] ~~.~~\label{ACT}
\end{equation}
This form is advantageous for numerical simulations, since one can
use the Pauli matrix representation for the $SU(2)$ matrices. 
It was found in \cite{bha} that this action has a first 
order bulk transition at $\b \sim 2.5$.  We have
checked that the two actions above give identical results, and then used
Eq.(\ref{ACT}) for our simulations. Another action 
that we used is the Halliday-Schwimmer action \cite{hal}
\begin{equation}
S={\v \over 2} \sum_p \sigma_p {\rm Tr}_f ~ U_p ~~.~~\label{HAL}
\end{equation}
Here $U_p$ is defined as before, but the link variables $U_\mu (x) \in
SU(2)$,
and $\sigma_p = \pm 1$. Besides the integration over the link variables, 
the partition function in this case also contains a summation over
all possible configurations of the set $\lbrace \sigma_p \rbrace$, thus
ensuring that the action is blind to the $Z(2)$ center symmetry of $SU(2)$. 
It is thus as good as Eq.(\ref{ADJ}) for exploring the role of $Z(2)$ in
deconfinement transition. It was found in \cite{hal} that the action 
(\ref{HAL}) shows a first order bulk phase transition at $\v \sim
4.5$. The chief advantage of this action is that both the link variables
$U_\mu$ and the plaquette variables $\sigma_p$ can be updated using 
heat-bath algorithms
\cite{cre}. We have used it for both qualitative studies of nature of
the high temperature phase in the next section and for quantitative 
investigations in later sections, where substantial computation was 
necessary.

One of the observables which we used to monitor 
the phase transitions is the adjoint plaquette variable $P$, defined as the 
average of $\frac{1} {3} {\rm Tr}_a U_p$ over all plaquettes for actions 
(\ref{ADJ}) and (\ref{ACT}); for the action (\ref{HAL}), $P$ is defined as 
the average of $\sigma_p{\rm Tr}_f U_p$  over all plaquettes.
The order parameter of the deconfinement transition in $SU(2)$ gauge theory, 
$\lf$, where $L_f$ is given by 
\begin{equation}
L_f (\vec r) = {\rm Tr}_f ~ \prod_{i=1}^{N_t} U_t (\vec r, i)~~,~~
\end{equation}
is identically zero for the actions (\ref{ACT}) and (\ref{HAL}) due to
their local $Z(2)$ symmetry.  Its natural analogue
for the $SO(3)$ theory is  $\l$, the average over all spatial sites of the 
adjoint Polyakov loop, defined by 
\begin{equation}
L_a (\vec r) = {\rm Tr}_a ~ \prod_{i=1}^{N_t} U_t (\vec r, i)~~. \label{POL}
\end{equation}
Note that $\l$, unlike $\lf$, is not an order parameter, as it
is not constrained to be zero in the confined phase. Since $\l$  and
$\lf$ can be thought of as measures of free energy of a fundamental
and an adjoint quark, respectively, their different behavior in the
confined phase is related to the fact that an adjoint
quark in the confined phase can be screened by gluons created from
the vacuum, while a fundamental quark cannot. For the same reason, an
adjoint Wilson loop is not supposed to exhibit the area law. However, 
creation of gluon pairs from vacuum costs a considerable amount of
energy as glueballs are heavy. It may therefore be favorable for adjoint 
quarks also to have a string between them, at least when they are not
too far separated. Intermediate size adjoint Wilson loops were found \cite{ber}
to show an area law for $SU(2)$ gauge theory, giving a string tension that 
is $\sim$2 times as large as the fundamental string tension. 
Furthermore, the behavior of the adjoint Polyakov loop across the $SU(2)$ 
deconfinement transition was found to be qualitatively similar to that of the 
fundamental Polyakov loop: $\l ~ \sim 0$ (for $8^3 \times 2$ and $\ea$
lattices) till the deconfinement transition, where it acquires a nonzero
value \cite{dam}. 
The jump in $\l$ (and also in even
higher representation Polyakov loops) is surprisingly similar to that
in $\lf$. This is believed to be related to opening of mass gap across
deconfinement: below the deconfinement transition, adjoint quark can exist
only by forming a bound state with gluon, which costs a lot of energy
and leads to a small expectation value for $\l$. 

	For the same reason, $\l$ can be expected to show a sharp change
at the deconfinement transition for $SO(3)$ gauge theory also and can,
therefore, serve as a good indicator of deconfinement transition
in $SO(3)$ gauge theory. 	The behavior of $\l$ in finite temperature
$SO(3)$ gauge theory was studied numerically in Ref. \cite{sri}. 
Using a $7^3 \times 3$ lattice and the action (\ref{ACT}), it was found
that $\l$ was consistent with zero till $\a \sim 2.5$, after which it became
nonzero, indicating a deconfinement transition around this value of $\a$.

\section{\label {sc3} The High Temperature Phase}

An unexpected and curious result of Ref. \cite{sri} was that after
becoming nonzero in the high temperature phase, $\l$ 
settles into either a positive value ($\to 3$ as $\a \to \infty$), or 
a negative value ($\to -1$ as $\a \to \infty$), the average value of the
action being the same for both the states. 
In \cite{sri} the negative $\l$ state was
interpreted as the manifestation of another zero temperature confined
phase. Since its negative value is inconsistent with its being the
exponential of the free energy of an adjoint quark,
it was conjectured that the negative value is 
a finite volume effect and that it should go to zero on bigger
lattices.  

We have carried out a number of tests in order to understand
the nature of the negative $\l$ state. First, it was checked that the
appearance of this phase is not due to any algorithmic problem, by
checking that it appears irrespective of whether one uses action
(\ref{ADJ}), (\ref{ACT}) or (\ref{HAL}). Since one uses explicitly
$SO(3)$ symmetric multiplication table for the first of these actions and
a heat-bath for the third, any doubts of the negative $\l$-phase
being an artifact of the $SU(2)$-based algorithm vanished, when it was
observed for all the three actions for the corresponding deconfined phases.
Indeed, unlike Ref. \cite{sri} or action (\ref{ACT}), where only the hot
starts in the deconfined phase lead to it, the heat-bath algorithm for
the action (\ref{HAL}) yielded it from even cold starts in the
deconfined phase.  Next, we checked whether the negative
value is a finite size effect  by simulating the theory on $N_s^3
\times 3$ lattices with $N_s$ ranging from 7 to 18. Our results are
presented in Table \ref{laval}. They indicate that the value of $\l$ is
quite stable against change in spatial lattice size. Looking at the
trend for $N_s = 9$ to $N_s = 18$ in Table \ref{laval}, one can 
estimate the value of $\l$ to be $\sim -0.64(1)$ on an 
$\infty^3 \times 3 $ lattice.

\vskip3cm
\begin{table}
\caption{$\l$ in the negative state for $\a = 3.5$ on 
$N_s^3 \times 3$ lattices with $N_s$ ranging from 7 to 18.}
\vskip2mm
\begin{center}
\begin{tabular}{|c||c|c|c|c|c|}
\hline
$N_s$ &7 &9 &12 &15 &18 \\
\hline
$\l$ &-0.656(1) &-0.642(3) &-0.643(3) &-0.639(3) &-0.641(4) \\
\hline
\end{tabular}
\end{center}
\label{laval}\end{table} 

The constancy of $\l$ in the negative phase above suggests it to be a
genuine state on an $\infty^3 \times N_t$ lattice. Just as the negative
$\lf$-phase of the $SU(2)$ theory is physically the same as the positive
$\lf$-phase, the negative $\l$-phase could be similar to the positive
$\l$-phase. A way to test this possibility is to introduce a polarizing 
``magnetic field'' by adding a term $h \sum_{\vec x} L_a(\vec x)$ to
the action (\ref{ACT}).  As shown in Fig. 1, the average plaquette 
$\langle P \rangle $ on a $7^3 \times 3$ lattice is not affected 
strongly by this term either below the transition ($\a=2.3$) or above 
the transition ($\a=3.5$).  However, $\l$ is.  Irrespective of the 
starting configuration, it always converges to a unique value 
whose sign is determined by that of $h$,
whereas for $h=0$ only some hot starts settled to negative $\l$.

\begin{figure}[htbp]
\begin{center}
\epsfig{height=8cm,width=8cm,file=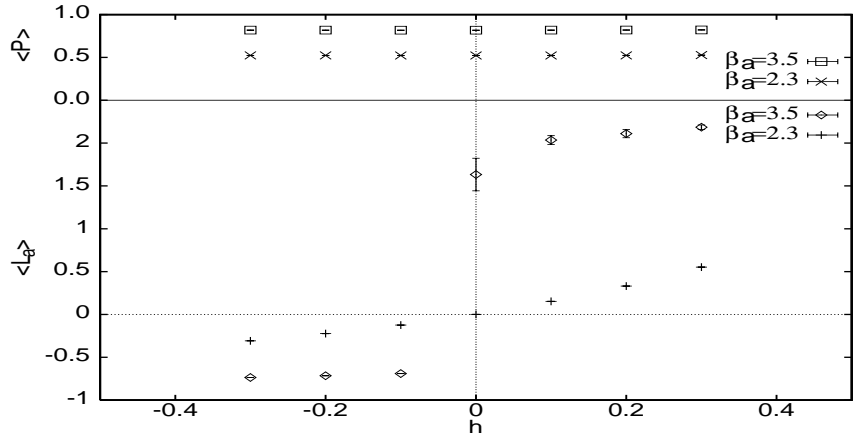}
\caption{$\langle P \rangle$ and 
$\l$ for a $7^3 \times 3$ lattice, for $\a = 2.3$
and 3.5, in the presence of a magnetic field $h$.} 
\end{center}\end{figure}

Defining $\l$ as 
\begin{equation}
\l = \lim_{h \to 0^+} \lim_{V \to \infty} {\partial \over \partial
h} \ln Z(h)~~, \label{MAG}
\end{equation}
in analogy with spin models, where one looks for a spontaneous breakdown of a
symmetry in this way, one also gets a positive $\l$ always.  This too is
similar to the $SU(2)$ case, except that the normalizations of $\l$
coming from the two different phases are different here, being 3 and 1,
respectively.  This suggests strongly that the high temperature phase of
the $SO(3)$ gauge theory also manifests itself in two ways corresponding
to positive and negative $\l$.  Also note in Fig. 1 that the same
definition of $\l$ yields a value consistent with zero below the phase
transition.

A further test of the similarity of the physics in these two phases is 
provided by the correlation functions in these phases. If the phases
are indeed physically similar, they ought to have the same correlation
lengths, and therefore, the same correlation functions apart from
normalizations. In fact, it has been argued \cite{smi} that even for 
the $SU(2)$ theory, the true order
parameter is the two point correlation function of $L_f$.
The behavior of the two point correlator is known to be quite different 
in the confined and the deconfined phases. In the limit of infinite
separation, it goes to zero in the confined phase and a constant in the
deconfined phase. In Fig. 2 we show $\Gamma(r)$ on an $\ea$ lattice, 
defined by 
\begin{equation}
\Gamma(r) = \sum_i \sum_{\vec x} \langle L_a(\vec x + r e_i) ~  L_a(\vec
x) \rangle~~,~~
\end{equation}
as a function of $r$ in lattice spacing units 
for $\a = 2.3$ and for the positive and negative 
$\l$ states at $\a = 2.6$ and 3.5. 
At each $\beta$ value $5 \times 10^6$ iterations were made.
The errors were calculated by dividing the measurements
into blocks of 5000 each. It was checked that
altering the bin-size does not change the error.
For $\a=2.3$, the $r$=4 point and the errors for 
the $r$=3 point are not shown, as the former has a negative central value, 
being consistent with zero within error, and the latter 
are of the order of the mean itself.
One clearly sees that (i) below the transition at $\a=2.3$, the
correlator vanishes rapidly with $r$, (ii) it approaches a constant above
the phase transition and (iii) the constant is bigger for larger $\a$ and
bigger in the positive $\l$-phase for the same $\a$.  

\begin{figure}[htbp]\begin{center}
\epsfig{height=8cm,width=8cm,file=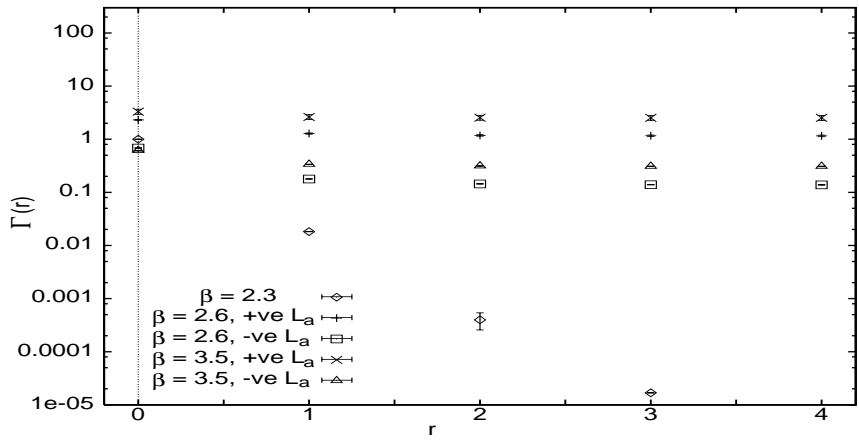}
\caption{The two-point correlator $\Gamma(r)$ plotted versus $r$
on a $\ea$ lattice for $\a = 2.3$, 2.6 and 3.5}
\end{center}\end{figure}

Finally, one can extract correlation lengths or mass gaps from these
correlation functions on sufficiently large lattices.
The mass gap can be obtained directly from the connected parts of the
correlator above or from their zero momentum projected versions.
Our intention here is only to compare the behavior
of the correlators in the low $\beta$-phase with that of the
correlators in the two states of the high $\beta$-phase.
Consequently, a small lattice should suffice as well; the mass gap so
obtained will be influenced by higher states which should,
however, be expected to be similar in the two high temperature phases.
It was found that due to large fluctuations in $L_a$, the
signals for connected part of the correlation function were difficult
to extract. However, the signal improved considerably by looking at
$\Gamma(r-1) - \Gamma(r)$, as shown in Fig. 3
(errors for $r$ = 4 are of the size of the correlation function itself and
are not shown for clarity). As expected for states with same physics,
the positive and negative $\l$ states corresponding to both $\a = 2.6$ and 3.5
have a similar mass gap, which does not change significantly
as one increases $\a = 2.6$ to 3.5. The mass gap is, however, 
considerably different for $\a = 2.3$. Interestingly, this picture too
matches well with the knowledge from $SU(2)$ gauge theory, where it has
been found that above the deconfinement transition, the mass gap
changes very little with coupling \cite{hel}.

\begin{figure}[htbp]\begin{center}
\epsfig{height=8cm,width=8cm,file=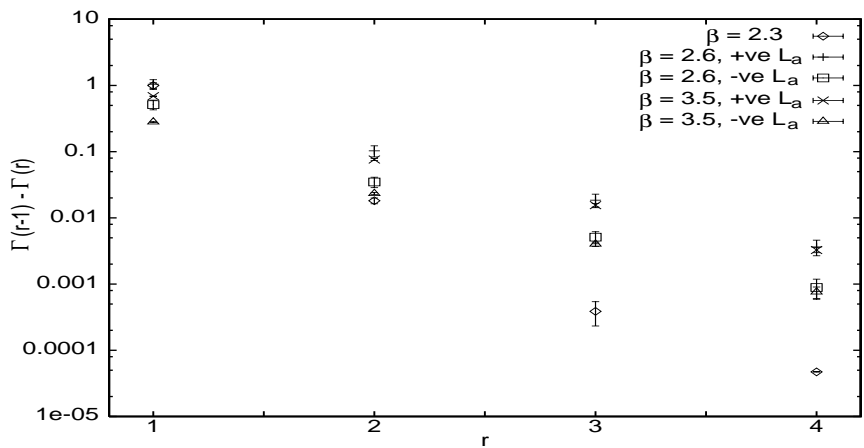}
\caption{$\Gamma(r-1) - \Gamma(r)$ vs r, for $\a = 2.3, 2.6, 3.5$.}
\end{center}\end{figure}

Essentially the same conclusions emerge from the zero momentum
correlators shown in Fig. 4.  Defining the zero momentum projection by
averaging the adjoint Polyakov loop over planes,
\begin{equation}
L_{ap}(x) = \sum_{y,z} L_a (x,y,z)~~,~~
\end{equation}
one defines its correlator in the usual way:
\begin{equation}
\Gamma_p(r) = \sum_{x} \langle L_{ap} (x+r) ~ L_{ap} (x) \rangle ~~.
\end{equation}
As is well known, a transfer matrix approach allows one to define the
mass gap from the connected parts of these correlators and again we
consider $\Gamma_p(r-1) - \Gamma_p(r)$ to reduce fluctuations.

\begin{figure}[htbp]\begin{center}
\epsfig{height=8cm,width=8cm,file=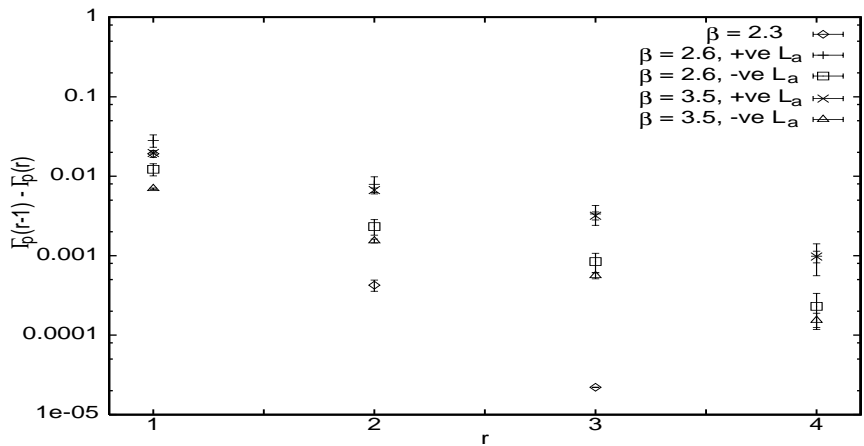}
\caption{Variation of the subtracted plane-plane correlator,
$\Gamma_p(r-1) - \Gamma_p(r)$ with $r$. The values of couplings are 2.3, 2.6,
3.5.}
\end{center}\end{figure}
 
In summary,
the effect of the external field $h$ on the two phases of $\l$ above the
transition and the behavior of the correlation functions in these
phases suggest strongly that they are physically the same phases.  Together 
with the corresponding results for the phase below the transition, they
further suggest that the phase transition is a deconfining one and the high 
temperature phase appears either as a positive or equivalently as a
negative $\l$-phase.

\section{\label{sc4} Order of the Transition}

	In order to determine the order of the transition, simulations
were made on $\fa$, $\sa$ and $\ea$ lattices with the action
(\ref{HAL}) and usual finite size scaling techniques were employed
\footnote[1]  
{ Exploratory studies were also done for action (\ref{ACT}) to check that 
they give similar results.  Some of these results can be found in \cite{sau}.
The only change in this case is that the transition occurs
at $\a \sim 2.5$ \cite{sri,sau}, but it displays the same features as
discussed in this section for action (\ref{HAL}).}.

Long lived metastable states were observed on all lattices near the transition 
region, signalling a possible first order transition.
Runtime evolutions of the plaquette $P$ and the Polyakov loop $L_a$
from different starting configurations, averaged over bins of 50 iterations, 
are presented in Fig. 5 for the $\ea$ lattice. Runs on smaller lattices, not
shown here, show more tunnellings and larger fluctuations in the
positive $L_a$-phase, but are otherwise similar in character.
The $L_a$ tunnels between 
all the three states, two of which correspond to the same value of the 
action, but different signs of $L_a$.  
The transition point was estimated by demanding equal probability in
the two phases for the action for these metastable states and error on
it was estimated by observing a lack of tunnelling.
For the $\fa$, $\sa$ and $\ea$ lattices the transition points are at 
$\beta_{vc} = 4.43 \pm 0.02$, $4.45 \pm 0.01$ and $4.45 \pm 0.01$,
respectively.

\begin{figure}[htbp]\begin{center}
\epsfig{height=6cm,width=5cm,file=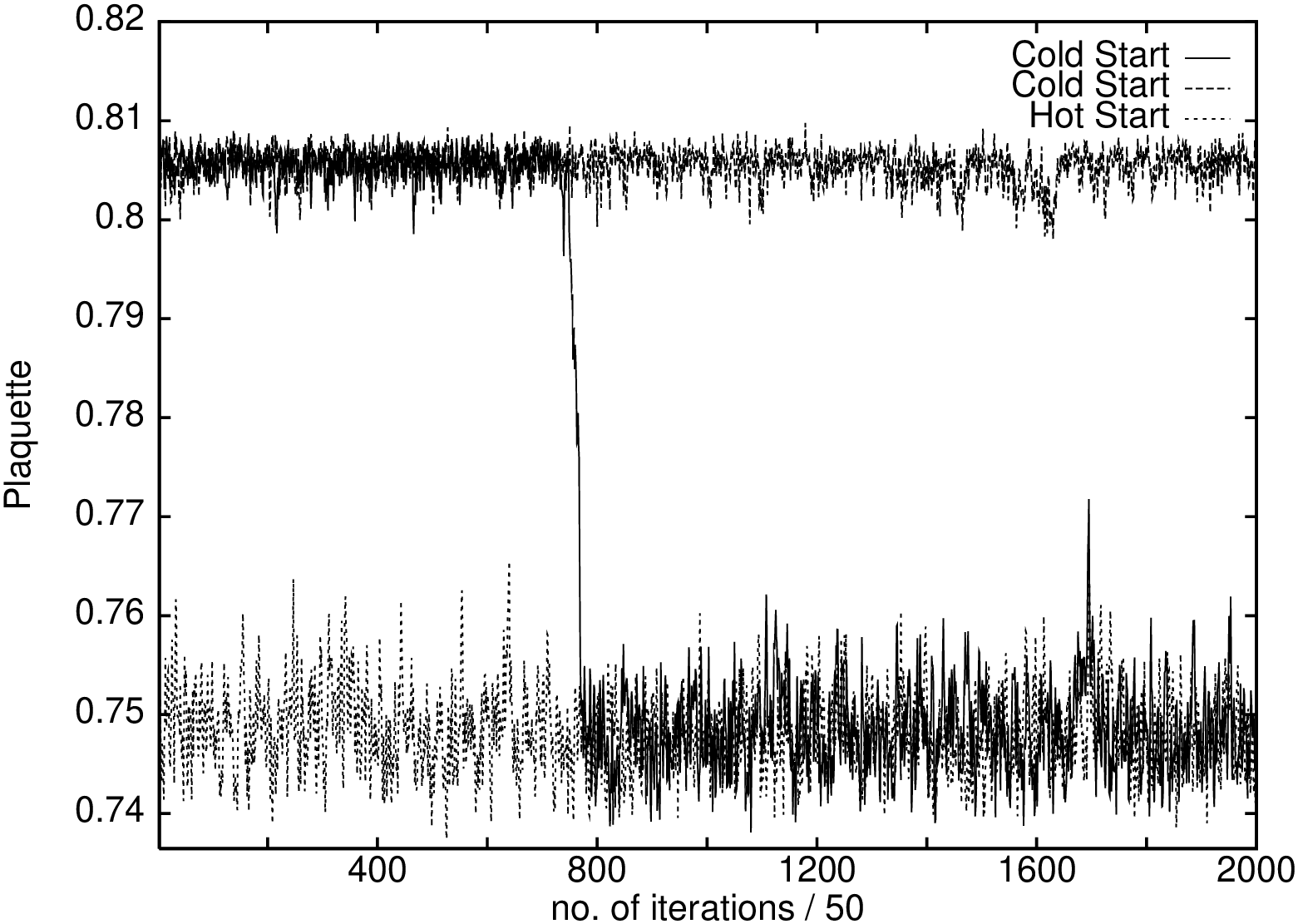}
\epsfig{height=6cm,width=5cm,file=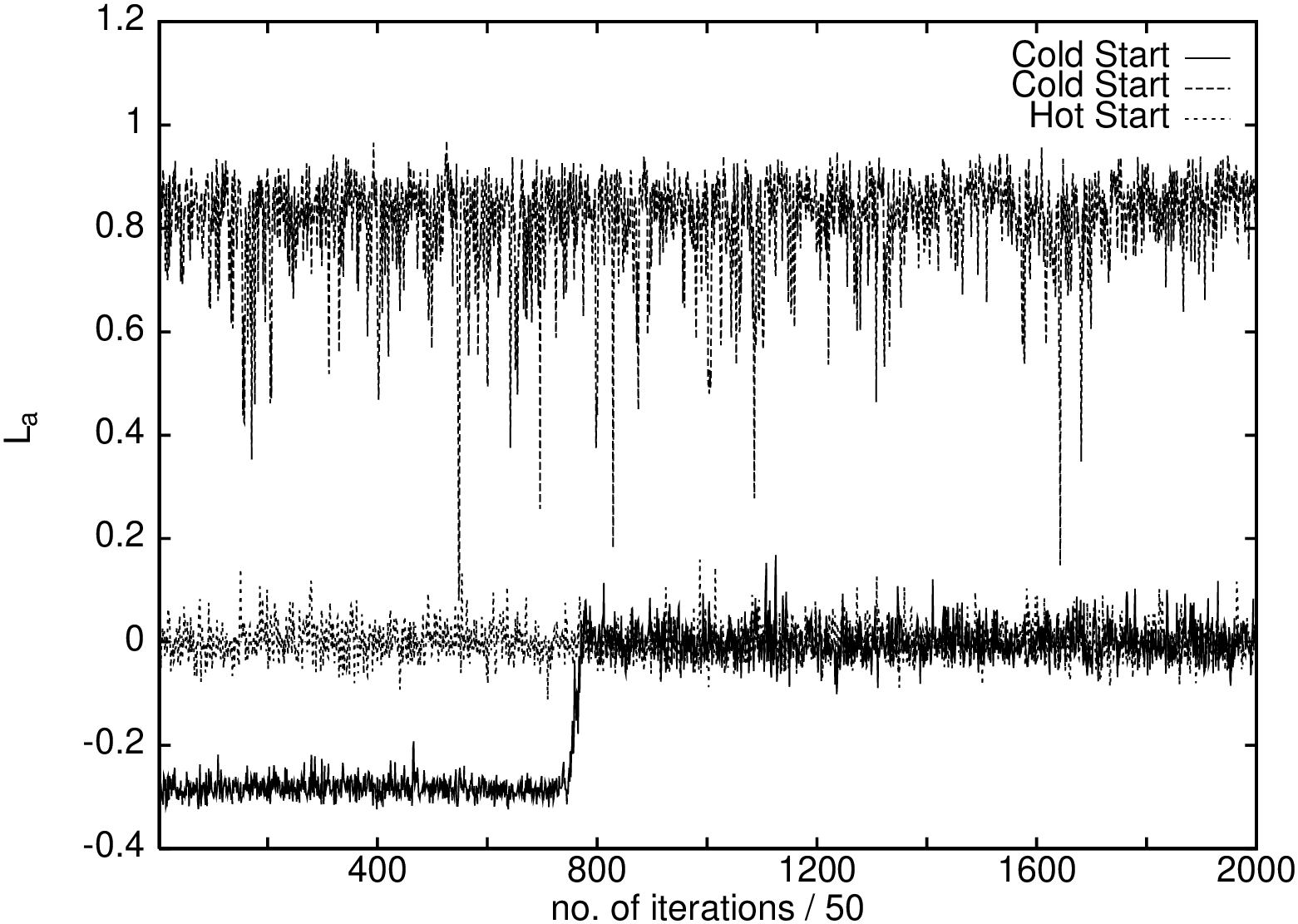}
\caption{a) Runtime evolution of plaquette for $\ea$ 
lattice. The values have been binned
over 50 iterations. b)Same for $\l$.}
\end{center}\end{figure}

\begin{table}
\caption{The discontinuities in the plaquette $\langle P \rangle $ and 
$\l$ at the transition temperature on $N_s^3 \times N_t$ lattices. 
The last two columns list the differences of $\l$ in the two high
temperature phases with that in the low temperature phase. The errors
correspond to the bin size used.}
\vskip2mm
\begin{center}
\begin{tabular}{|c|c|c|c|c|c|}
\hline
$N_t$ &$N_s$ &$\beta_{vc}$ &$\Delta P$ &$\Delta L_{a_+}$ &$\Delta 
L_{a_-}$ \\
\hline
\hline
4 &4 &4.43 &.0630(30) &.92(6) &- \\
\hline
4 &6 &4.45 &.0575(30) &.86(4) &.26(2) \\
\hline
4 &8 &4.45 &.0575(30) &.87(4) &.28(4) \\
\hline
\hline
6 &6 &4.45 &.0575(30) &.42(4) &.13(4) \\
\hline
8 &8 &4.45 &.0575(30) &.20(4) &.04(4) \\
\hline
\end{tabular}
\end{center}
\label{finsize}\end{table} 

In order to confirm the above indications of a first order transition in a 
more quantitative study, the distributions of the plaquette and $\l$ were
analyzed. Figure 6 displays the distributions of the plaquette
variable on the lattices studied from the runs
made at the critical couplings, but from different starts. 
We performed about 100-400 K heat-bath sweeps depending on the size of the
lattice.  There is a clear two-peak structure in the distributions. 
While the position of one of the peaks shift slightly in going from
$\fa$ to $\sa$ lattice, no shift is seen in going
from $\sa$ to $\ea$ lattice. Assuming the peak positions to correspond
to the expectation values in $N_s \to \infty$ limit, the estimates of the 
discontinuities in the plaquette are presented in Table \ref{finsize}.
Clearly, the 
plaquette discontinuity remains constant with increasing lattice size. As
seen from Fig. 6, the valley between the peaks becomes steeper
with increasing spatial size for the lattice, signalling again a first
order
transition. The corresponding distributions for the Polyakov loop $L_a$ are 
presented in Fig. 7, and the estimates of the
discontinuity for both the positive and negative $L_a$ phases are also given 
in Table \ref{finsize}. While the frequent tunnelling smoothens the
peak structure for 
the $\fa$ lattice considerably, a clear three-peak structure is seen for 
both the $\sa$ and the $\ea$ lattices. 
Once again the peak positions are seen not to shift and the valley between 
peaks is seen to become steeper with increasing lattice size, pointing to 
a finite discontinuity in the infinite volume limit and
a first order transition.  It is also interesting to note that the
peak for the confined phase is almost precisely at zero.
As argued in Sec. \ref{sc2}, one expects to see a linearly rising 
potential between static adjoint quarks in the confined phase of the
$SO(3)$ theory for intermediate distances. 
The leading order strong coupling contribution to $\langle L_a(\vec r)
L_a(\vec 0) \rangle$, $~\propto ~\exp ~(- V(r,T)/T)$, is $\b ^ {r N_t}$
for a set of plaquettes spread between the loops, and $\b ^ {8
N_t}$ for tubes around the loops. Thus if one is still in the leading
order strong coupling regime at $\v=4.45$, one expects to see a linearly 
rising potential for lattice distances up to 8. This may explain $\l$ =0
in the confined phase.  We have, however, checked that even on a $16^3
\times 4$ lattice, it continues to remain zero and the histograms in
Figs. 6 and 7 do not shift at all, but become sharper and narrower.

\begin{figure}[htbp]\begin{center}
\epsfig{height=8cm,width=8cm,file=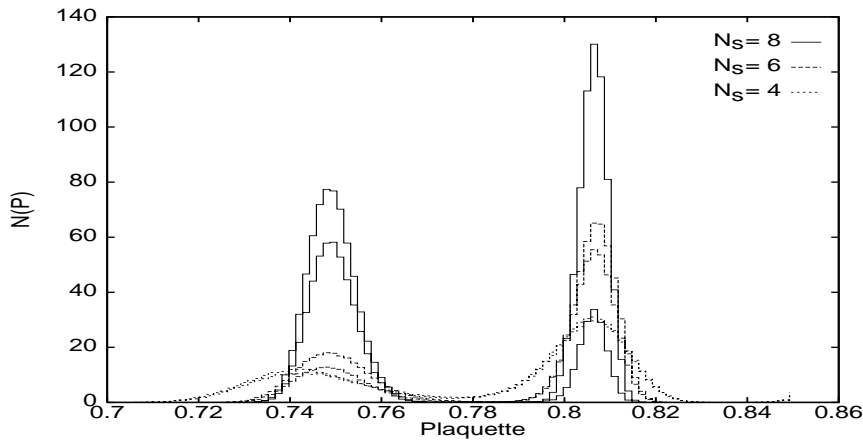}
\caption{The distribution of the plaquette on $N_s^3 \times 4$
lattices at the critical couplings given in Table {\protect \ref{finsize}}.}
\end{center}\end{figure}

\begin{figure}[htbp]\begin{center}
\epsfig{height=8cm,width=8cm,file=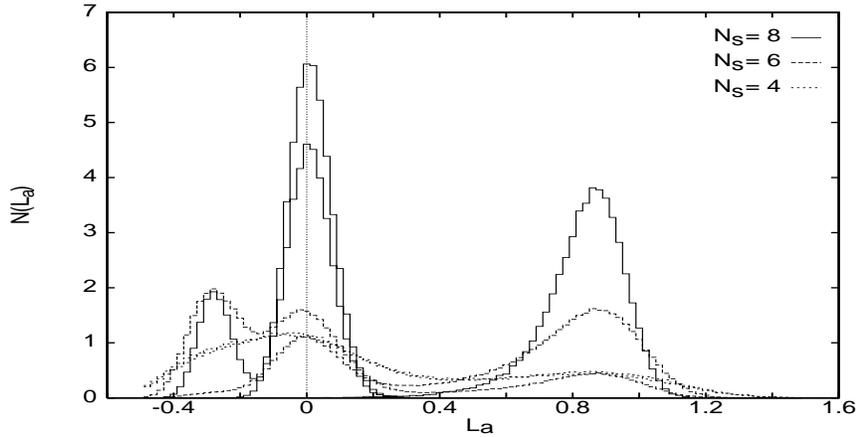}
\caption{Distribution of $L_a$ on $N_s^3 \times 4$ lattices for 
couplings of Table {\protect \ref{finsize}}.} 
\end{center}\end{figure}

\section{\label{sc5} Nature of the Transition}

	As mentioned in the Introduction, $SO(3)$ lattice gauge theory
is supposed to have a bulk transition \cite{bha,hal}, while we argued 
above that the only transition seen on $N_t= 4$ lattices is more
appropriately identified as the deconfinement phase transition at high
temperatures.  In this section we attempt to address the issue of bulk
transition.

\begin{table}
\caption{The critical coupling $\bc$ as a function of the lattice
size in the time direction, for the actions ({\protect\ref{ACT}}) and
({\protect\ref{HAL}}). Also presented are the corresponding values of $\b_c$ 
for pure $SU(2)$ gauge theory with the action ({\protect\ref{WIL}}) (taken
from ({\protect\cite{fin2}})).}
\vskip2mm
\begin{center}
\begin{tabular}{|c|c|c||c|}
\hline
 &$SO(3)$ &$SO(3)$ &$SU(2)$ \\
\hline
$N_t$ &action ({\protect\ref{HAL}}) &action ({\protect\ref{ACT}})
&action ({\protect\ref{WIL}}) \\
\hline
\hline
2 &4.156(10) &2.415(5) &2.1768(30) \\       
\hline
4 &4.43(2) &2.53(1) &2.2986(6) \\
\hline
6 &4.45(2) &2.52(2) &2.4265(30) \\
\hline
8 &4.45(2) &2.52(2) &2.5115(40) \\
\hline
\end{tabular}
\end{center}
\label{coup}\end{table} 

Since the deconfinement temperature is a physical quantity (in the
hypothetical world of 2 colors and only gluons), it is expected 
to remain constant under a change of $N_t$: 
$T_c = 1 / N_t a(\bc)$ implies that a change in $N_t$ should  merely
change $\beta_c$ and push it to larger $\b$ as $N_t$ is increased.
In order to check this, we studied the theory
on $\tw$, $\fs$, $\xs$ and $\es$ lattices.
On all these lattices, only one transition point were found, where
both the plaquette and $\l$ show a discontinuity. 
The critical couplings for $N_t = 2, 4, 6, 8$, extracted from the
runs made on the lattices above, are presented
in Table \ref{coup}. We have also included in the table the corresponding
critical couplings for action (\ref{ACT}) from our own work
and those for the deconfinement transition in $SU(2)$
gauge theory, taken from Ref. \cite{fin2}.

	It is found from Table \ref{coup} that as one goes from $N_t = 2$ to
$N_t = 4$, there is a clear shift in $\b_c$ in all the three cases.
This behavior is consistent with the deconfinement scenario. 
However, no perceptible change in $\b_c$ was found for either of the
actions for $SO(3)$ in going from $N_t = 4$ to 6 and 8.
This is in sharp contrast to the $SU(2)$ case, and is also unexpected 
for a deconfinement transition; the behavior, however, is similar to
that of the transition seen in Ref. \cite{gav2} for $SU(2)$ gauge 
theory with action (\ref{BHA}). The distributions for the plaquette $P$
are exhibited in Fig. 8. They again suggest a first order phase
transition and the estimated discontinuity in plaquette
is listed in Table \ref{finsize}. One sees that it remains constant as one increases
$N_t$. We have also looked at the corresponding 
distributions of $\l$ for these lattices. In spite of the noisy signals
due to small spatial sizes, a three-peak structure could still be ascertained
in all of them. Table \ref{finsize} lists the corresponding discontinuities for $\l$. 
It should be noted that $\lf$ at the transition point decreases with
$N_t$ for both $SU(2)$ and $SU(3)$ theories. The decrease in the
discontinuities in $\l$ in Table \ref{finsize} are for similar reasons.

\begin{figure}[htbp]\begin{center}
\epsfig{height=8cm,width=8cm,file=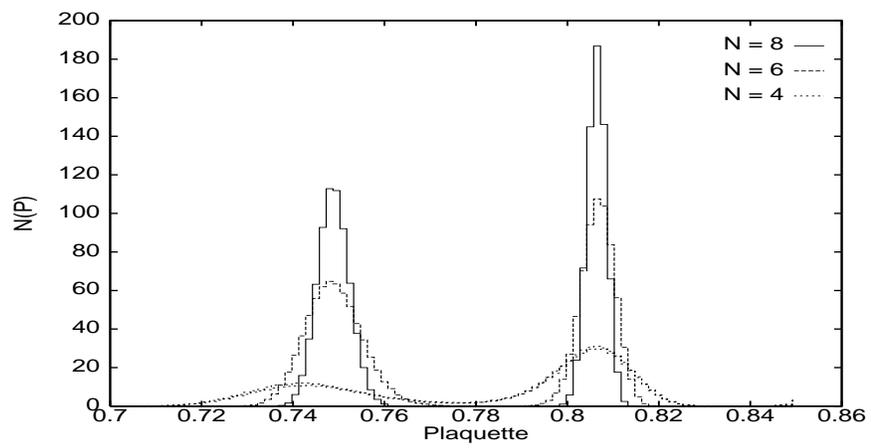}
\caption{Distributions of the plaquette on $N^4$
lattices at the critical couplings given in Table {\protect \ref{finsize}}.}
\end{center}\end{figure}

\section{\label{sc6} Summary and Discussion}
The study of phase transitions in $SO(3)$ gauge theory is important for
understanding both the interplay of the bulk and the deconfinement transition
and the nature of its deconfinement as it has no center symmetry.
The theory was studied in Ref. \cite{sri} on a $7^3 \times 3$ lattice and a 
deconfining phase transition at $\beta_{ac} \sim 2.5$ was reported. 
For $\a > \beta_{ac}$, 
$\l$ was found to take either a positive value or a negative value. 
The positive $\l$ state was taken to correspond to the high temperature
deconfined phase, while the negative $\l$ state was interpreted as
being another zero temperature confined phase. 

	Our simulations with a variety of actions confirmed the results of 
Ref. \cite{sri}.  In particular, the negative $\l$-state was present in all 
of them.  However, using a ``magnetic field'' term to polarize, we found
 unique $\l$ state depending on the sign of the field.
The correlation function measurements in both the phases of positive and 
negative $\l$ also indicated that the two states are physically
identical. Both of these correspond to the high temperature deconfined phase
of $SO(3)$ gauge theory, as the correlators approached a nonzero constant in
the large separation limit, while below the transition a confined phase
was indicated by their exponential drop to zero.

	By studying the system on lattices of different sizes and different
aspect ratios, it was
established that there is only one phase transition for this theory,
which is of first order. In addition to the average action, 
the adjoint Polyakov loop also showed a jump across
the transition. Its vanishing until the transition point
further supports the interpretation of a deconfining transition.
The correlation lengths below and above $\bc$ behave similar to the
correlation lengths near the deconfinement transition of $SU(N)$ 
gauge theories.
If the transition is accepted to be a deconfinement phase transition,
then its first order nature is as puzzling as the observations of 
Ref. \cite{gav2}, since the $SU(2)$
gauge theory is known to have an Ising model-like second order 
phase transition.

On increasing $N_t$ from 4 to 6 to 8, the transition point did
not move. While investigations on still larger lattices will be required
to conclude firmly, this observed behavior does go
against the usual expectations of a deconfinement
transition. Since we did not find any other transition, one might be 
inclined to accept either
a coincident deconfinement transition or a total lack of a deconfinement
phase transition for $SO(3)$ gauge theory.  If it is the former then it 
is remarkably similar to the results for the mixed action \cite{gav1,gav2},
where too a shift in $\bc$ was observed only in going from $N_t$=2 to 4
for large $\a$, but no further shift occurred in changing $N_t$ up to 8.
Very large lattices are therefore necessary to
see the similarity of $SO(3)$ and $SU(2)$ theories at finite temperature,
if at all. The second alternative is incompatible with the behavior of 
$\l$ and its correlation function across the phase transition.
It is also clearly in disagreement with the naive expectations of 
purely gluonic confinement for $SO(3)$ gauge theory. 

\section{Acknowledgments}
	We thank Dr. Srinath Cheluvaraja and Dr. Sourendu Gupta for
many helpful discussions. It is a pleasure for one of us (R. V. G.)
to thank Professors Frithjof Karsch and Helmut Satz and the staff at the
Zentrum f\"ur interdisziplin\"are Forschung, Universit\"at Bielefeld
for their kind hospitality.

\end{document}